\documentclass[12pt]{article}
\usepackage{heck}
\usepackage{amsmath,amssymb}
\usepackage{color}
\definecolor{myblue}{rgb}{.97,.97,1}
\usepackage{empheq}
\usepackage{CJK}
\usepackage{hyperref}

\newlength\mytemplen
\newsavebox\mytempbox

\makeatletter
\newcommand\mybluebox{%
    \@ifnextchar[
       {\@mybluebox}%
       {\@mybluebox[0pt]}}

\def\@mybluebox[#1]{%
    \@ifnextchar[
       {\@@mybluebox[#1]}%
       {\@@mybluebox[#1][0pt]}}

\def\@@mybluebox[#1][#2]#3{
    \sbox\mytempbox{#3}%
    \mytemplen\ht\mytempbox
    \advance\mytemplen #1\relax
    \ht\mytempbox\mytemplen
    \mytemplen\dp\mytempbox
    \advance\mytemplen #2\relax
    \dp\mytempbox\mytemplen
    \colorbox{myblue}{\hspace{1em}\usebox{\mytempbox}\hspace{1em}}}
\makeatother    

\newcommand{\bZ}{\ensuremath{\mathbb{Z}}}

\newcommand{\beq}{\begin{equation}\begin{aligned}}
\newcommand{\eeq}{\end{aligned}\end{equation}}

\def\frakp{\ensuremath{\mathfrak{p}}}

\newcommand{\fl}{\mathfrak{l}}
\newcommand{\fn}{\mathfrak{n}}

\begin{document}
\begin{CJK*}{UTF8}{min}

\title{Comments on Determinant Formulas\\
for General CFTs}

\authors{\centerline{Masahito Yamazaki (山崎雅人)}}

\date{January, 2016}

\institution{IPMU}{\centerline{Kavli IPMU (WPI), University of Tokyo, Kashiwa, Chiba 277-8583, Japan}}
\institution{IAS}{\centerline{and School of Natural Sciences, Institute for Advanced Study, Princeton NJ 08540, USA}}

\abstract{
We point out that the determinant formula for a parabolic Verma module plays a key role in the study of (super)conformal field theories and in particular their (super)conformal blocks. The determinant formula is known from the old work of Jantzen for bosonic conformal algebras, and we present a conjecture for superconformal algebras. The application of the formula includes derivation of the unitary bound and recursion relations for conformal blocks.}


\maketitle 
\end{CJK*}

\tableofcontents


\section{Introduction}

In the standard textbook discussion of two-dimensional conformal field theories (CFTs),
one crucial ingredient is the celebrated Kac determinant formula \cite{Kac_LNP,Feigin_Fuks_1982}.
In light of recent exciting developments in conformal bootstrap \cite{Ferrara:1973yt,Polyakov:1974gs} in higher dimensions\footnote{See \cite{Rattazzi:2008pe, Rychkov:2009ij} for early pioneering papers, and references [5-26] of \cite{Penedones:2015aga} for a more complete list.},
a natural question is whether there is a generalization of the Kac determinant formula
to higher-dimensional CFTs, and if yes whether the formula is useful at all for the study of 
$D>2$ CFTs.

The goal of this small note is to point out that the there is indeed such a formula for a 
general semi-simple Lie algebra, and that it is relevant in the study of conformal blocks.
We also comment on possible generalizations to supersymmetric CFTs.
The utility of the determinant formulas (and more sophisticated Kazhdan-Lusztig theory \cite{KazhdanLusztig})
is illustrated further in a recent preprint \cite{Penedones:2015aga} with J.~Penedones and E.~Trevisani.

This short note grew out of the author's desire to 
bring higher-dimensional CFTs closer in spirit to the 
textbook treatment of 2d CFTs, where significant part of the structure is encoded in the 
representation theory of the Virasoro algebra.
We believe that there are a lot of rich mathematical structures yet to be uncovered
in the representation theory of parabolic Verma modules, and we would like to 
urge the representation-theory experts to take this subject seriously (if they have not done so already). 
For example,
very little seems to be known about explicit expressions of null states,\footnote{The null states give rise to 
conformal-invariant differential equations, and are identified with the so-called partially massless fields in the 
AdS dual, see e.g.~\cite{Eastwood:1987ki, Dolan:2001ih,Shaynkman:2004vu}.}
except for the three-dimensional CFT case worked out in \cite{Penedones:2015aga}.
As commented later in section \ref{sec.applications}, a better understanding of the representation theory of parabolic Verma modules
will be a crucial ingredient in the 
systematic study of general correlators (such as correlators of currents and stress-energy tensors)
for superconformal field theories in various dimensions and 
various amount of supersymmetry.

\section{Determinant Formula}

\subsection{Parabolic Verma Modules}

The representation theory of the conformal (or superconformal) algebra has a long history (see e.g.\ \cite{Mack:1975je,Dobrev:1985qv,Dobrev:1985vh,Dobrev:1985qz,Minwalla:1997ka,Dolan:2002zh} for an incomplete list). We here highlight the importance of parabolic Verma modules (also known as generalized Verma modules).
This module can be regarded as a higher-dimensional CFT counterpart of the 
well-known Verma module for the Virasoro algebra\footnote{Historically D.N.~Verma in his thesis considered a Verma module for a semisimple Lie algebra, before people started discussing Verma modules for the Virasoro algebras. Since the conformal group in higher dimensions is finite-dimensional, the structure of the Verma module for a higher-dimensional conformal group is on the one hand much simpler than in the two-dimensional conformal group 
involving an infinite-dimensional algebra. One the other hand, higher-dimensional case is more complicated in that we are interested in the generalized concept of the Verma module, namely a parabolic Verma module.}. While known to experts, these Verma modules deserve more attention 
from a wider spectrum of physicists interested in CFTs.

The conformal group in $D$ dimensions is given by $SO(D,2)$ in Lorentzian signature,
which group is generated by dilatation $D$, translations $P^{\mu}$, special conformal transformations $K^{\mu}$, and rotations $J^{\mu \nu}$,
with $\mu, \nu=1 \ldots D$. For our considerations (except the discussion of the unitarity bound towards the end of this paper)
the signature of the group is not important\footnote{Mathematically it is enough to have a complex form of $\mathfrak{g}$.}, and we instead consider the group $G=SO(D+2)$. We denote the associated semisimple Lie algebra by $\mathfrak{g}=\mathfrak{so}(D+2)$.


The conformal algebra $\mathfrak{g}$ has the following
decomposition
\beq
\mathfrak{g}=\mathfrak{n}_+\oplus \mathfrak{l} \oplus \mathfrak{n}_{-} \ ,
\label{gdecomp}
\eeq
where the subalgebras $\mathfrak{l}, \mathfrak{n}_{+}$ and $\mathfrak{n}_{-}$
are defined by
\beq
\mathfrak{l}:=\langle D, J_{\mu \nu} \rangle  \ , \quad
\mathfrak{n}_{+}: =\langle K_{\mu} \rangle  \ , \quad
\mathfrak{n}_{-}: =\langle P_{\mu} \rangle  \  .
\label{gdecomp2}
\eeq
The subalgebra $\mathfrak{l}=\mathfrak{u}(1)\oplus \mathfrak{so}(D)$, which is sometimes called the Levi subalgebra (hence the notation),
is a subalgebra which acts on the Hilbert space of 
the radially-quantized CFT.
The decomposition \eqref{gdecomp} satisfies
\beq
\left[\mathfrak{l}, \mathfrak{l} \right] \subset \mathfrak{l}  \ , \quad
\left[\mathfrak{l}, \mathfrak{n}_{\pm} \right] \subset \mathfrak{n}_{\pm}  \ , \quad
\left[\mathfrak{n}_{+}, \mathfrak{n}_{-} \right] \subset \mathfrak{l} \ .
\label{comm}
\eeq
This in particular means that the decomposition \eqref{gdecomp}
is preserved by an adjoint action of an element of $\mathfrak{l}$.
Of course, this is expected since $\mathfrak{l}$ is a subalgebra generated by rotations and the dilatation.

Given a decomposition \eqref{gdecomp}, a conformal primary is specified by a representation 
$V_{\lambda}$ of $\mathfrak{l}$, where $\lambda$ is a highest weight.\footnote{Or lowest weight in more physics-oriented conventions.} If we choose a basis $\{ |\Omega_{\lambda}^a \rangle\}_{a=1}^{\textrm{dim} V_{\lambda}}$ of this representation,
we have
\beq
 h  |\Omega_{\lambda}^a \rangle = \sum_b m_{\lambda}(h)^a{}_b |\Omega_{\lambda}^b \rangle
 \ , \quad \forall 
h\in \mathfrak{l} \ ,
\label{primary_constraint_0}
\eeq
where $m_{\lambda}$ is a matrix representation of $V_{\lambda}$.
Note that the highest weight $\lambda$ for the subalgebra $\mathfrak{l}$ in 
\eqref{gdecomp2} is given by a pair $(\Delta, \vec{l})$, with $\Delta$ specifying the 
operator dimension and the non-negative integers $\vec{l}$ specifying the spins of the
rotation group. The constraint \eqref{primary_constraint_0} then says that  $|\Omega_{\lambda}\rangle$ 
is a state with a specific operator dimension $\Delta$ (energy in radial quantization) and spins $\vec{l}$.

So far there is a representation only of $\mathfrak{l}$, and we wish to extend this to a representation of the full algebra $\mathfrak{g}$.
Now the crucial point is that  we need to impose the primary constraint:
\beq
 \mathfrak{n}_{+}  |\Omega_{\lambda}^a\rangle =0  \ .
\label{primary_constraint_1}
\eeq

Starting with the conformal primary $|\Omega_{\lambda}\rangle$ we can consider its descendants.
These are generated by the action of the subalgebra $\mathfrak{n}_{-}$,
which spans the parabolic Verma module:
\beq
M_{\mathfrak{p}}(\lambda): = \textrm{span} \left\{ g_1 g_2 \ldots g_k |\Omega_{\lambda}^a\rangle
\, \big| \, g_1, g_2, \ldots, g_k \in \mathfrak{n}_{-} \right\}  \ .
\eeq
(We will comment on the meaning of the index $\mathfrak{p}$ momentarily.)
In our case at hand, this is nothing but the conformal family of the primary $|\Omega_{\lambda}\rangle$:
\beq
M_{\mathfrak{p}}\left(\lambda=(\Delta, \vec{l})\right):= \textrm{span} \left\{ P^{\mu_1} P^{\mu_2} \ldots P^{\mu_k} \left|\Omega^a_{\lambda=(\Delta, \vec{l})}\right\rangle
\right\}  \ .
\eeq
We study this parabolic Verma module $M_{\mathfrak{p}}(\lambda)$ in the rest of this paper.

\bigskip
Here are some supplementary remarks.
First, let us explain the reason for the name ``parabolic'' (and the notation $M_{\mathfrak{p}}$). From the decomposition \eqref{gdecomp}
we can define a parabolic subalgebra $\mathfrak{p}$ by\footnote{This decomposition is known as the Levi decomposition of a parabolic subalgebra.}
\begin{align}
\mathfrak{p} := \mathfrak{n}_{+}\oplus \mathfrak{l} \;.
\end{align}
The primary constraints (\eqref{primary_constraint_0} and \eqref{primary_constraint_1}) then imply that $|\Omega_{\lambda} \rangle$ is a representation of 
$\mathfrak{p}$. Given such a representation $V_{\lambda}$ (whose highest weight we denote by $\lambda$), we can define the associated parabolic Verma module by\footnote{This 
is an example of an induced representation, and is sometimes denoted by $\textrm{Ind}^{\mathfrak{g}}_{\mathfrak{p}}(V_{\lambda})$.
}
\begin{align}
M_{\mathfrak{p}}(\lambda)=M_{\mathfrak{p}}(V_{\lambda}):=\mathfrak{U}(\mathfrak{g}) \otimes_{\mathfrak{U}(\mathfrak{p}) } V_{\lambda} \;,
\end{align}
where $\mathfrak{U}(\mathfrak{g})$ and $\mathfrak{U}(\mathfrak{p})$
are the universal enveloping algebras for $\mathfrak{g}$ and $\mathfrak{p}$, respectively,
and $V_{\lambda}$ here is naturally regarded as a representation of the universal enveloping algebra $\mathfrak{U}(\mathfrak{p})$.
We can verify that this general definition reduces to our previous definition of the parabolic Verma module 
$M_{\mathfrak{p}}(\lambda)$ for the conformal algebra.
Let us note that compared with the general discussion of parabolic Verma modules,
our discussion of conformal algebras is simpler, in that 
both $\mathfrak{n}_{+}$ and $\mathfrak{n}_{-}$ are Abelian, and 
that $V_{\lambda}$ is a trivial representation of $\mathfrak{n}_{+}$\footnote{In fact, by abuse of notation we are using the same symbol $\lambda$ for the highest weight, both for a representation of $\mathfrak{p}$ and for that of $\mathfrak{l}$.}.

Second, instead of the decomposition \eqref{gdecomp2} we can consider the 
triangular decomposition of the Lie algebra, with $\mathfrak{l}$ given by the Cartan subalgebra.
The parabolic Verma module then reduces to the ordinary Verma module,
and the parabolic subalgebra $\mathfrak{p}$ coincides with the Borel subalgebra.
This case is familiar from the representation theory of the Virasoro algebra,
in which case the decomposition \eqref{gdecomp}
reads
\beq
\mathfrak{n}_{+}=\langle L_{n>0} \rangle \ , \quad
\mathfrak{l}=\langle L_{n=0} \rangle \ , \quad
\mathfrak{n}_{-}=\langle L_{n<0} \rangle \ ,
\eeq
where $L_n \, (n\in \bZ)$ are the generators of the Virasoro algebra.\footnote{
In two dimensions, parabolic Verma modules have been studied in the context of 
logarithmic CFTs.
}

\bigskip
For physical considerations it is important that the parabolic Verma module $M_{\frakp}(\lambda)$ is equipped with an inner product (the so-called 
contravariant form).
This can be done specifying the conjugation in the sense of radial quantization (the so-called BPZ conjugation)
\beq
D^{\dagger}=D \;, \quad
(P_{\mu})^{\dagger}=K _{\mu} \;,  \quad
(J_{\mu \nu})^{\dagger}=J_{\mu \nu} \;,
\label{BPZ}
\eeq
and canonically normalizing the primary state $|\Omega_{\lambda}\rangle$.
More generally, we can define an inner product on $M_{\mathfrak{p}}(\lambda)$ by a 
suitable anti-involution $\sigma$ exchanging $\mathfrak{n}_{+}$ and $\mathfrak{n}_{-}$, and choosing an inner product to satisfy
$\langle g \cdot u  |v\rangle= \langle  u | \sigma(g)\cdot v\rangle$.

Once we have an inner product in $M_{\mathfrak{p}}(\lambda)$,
we can define the determinant.
While $M_{\mathfrak{p}}(\lambda)$ in itself is an infinite-dimensional representation (and has an ill-defined determinant),
$M_{\mathfrak{p}}(\lambda)$ naturally decomposes into a sum (an infinite sum) of
finite-dimensional representations (with highest weight $\mu$) of $\mathfrak{p}$ (or rather $\mathfrak{l}$ for our discussion of the conformal group)
\beq
M_{\mathfrak{p}}(\lambda)=
\displaystyle\bigoplus_{\mu} M_{\mathfrak{p}}(\lambda)^{\mu} \;.
\eeq 
Let us impose the condition that the inner product is zero between two states with different $\mu$'s.
This condition is satisfied for the the conjugation of \eqref{BPZ}.
The determinant then factorizes into a product of the determinant in each of these summands,
where the latter can be defined by choosing a basis $\{|v_1\rangle, |v_2\rangle, \ldots \}$ of $M_{\mathfrak{p}}(\lambda)_{\mu}$:
\beq
\det M_{\mathfrak{p}}(\lambda)^{\mu}=
\det_{i,j} (\langle v_i | v_j \rangle) \;.
\label{detM}
\eeq 
Note also that we are primarily interested in the zeros of the determinant, which are not affected by the choice of the 
basis in \eqref{detM}.

\subsection{Determinant Formula and Simplicity Criterion}

The paper by Jantzen \cite{Jantzen}, which seems to be little-known in the physics literature,
gives an explicit form of the determinant \eqref{detM} (see also \cite{Humphreys}).

The formula, stated for our parabolic Verma module $M_{\mathfrak{p}}(\lambda)$, reads \cite[Satz2]{Jantzen}:
\begin{empheq}[box={\mybluebox[7pt]}]{equation}
\textrm{det}\, M_{\mathfrak{p}}(\lambda)^{\mu}= 
\textrm{const.} \prod_{\beta\in {\Delta_{\fn}}} \prod_{n>0} 
\left(
\langle \lambda+\rho , \beta^{\vee}\rangle-n 
\right)^{\textrm{ch}\, M_{\mathfrak{p}}(\lambda-n \beta)_{\mu}} \ .
\label{bosonic_determinant}
\end{empheq}

Let us explain the symbols in this formula. First, the only important thing to know about the 
($\lambda$-dependent, but $\mu$-independent) constant in front of the determinant is that it is always non-zero.

Let us next explain the symbol $\Delta_{\fn}$. For the semisimple Lie algebra $\mathfrak{g}$ let us denote by $\Delta$ the set of simple roots, 
and by $\Delta^{+}$ the set of positive roots (with respect to a certain basis of $\Delta$). 
Let us denote by $\Pi_{\mathfrak{l}}$ the roots of $\Delta$ whose corresponding element of $\mathfrak{g}$ (in the Weyl-Cartan basis) are in the subalgebra $\mathfrak{l}$. We then define $\Delta_{\mathfrak{l}}:=\mathbb{Z} \Pi_{\mathfrak{l}} \cup \Delta$.\footnote{Note that $\mathfrak{l}$ is not semi-simple ($\mathfrak{l}=\mathfrak{so}(D)\oplus \mathfrak{so}(2)$), and  $\Delta_{\mathfrak{l}}$ defined this way 
contains the roots in $\mathfrak{g}$ corresponding to the Abelian part of $\mathfrak{l}$,
namely the root for the dilatation generator $D$. Note also that we choose the basis of $\Delta_{\fl}$ as a subset of the basis for $\Delta$.}
We further define the positive part by $\Delta_{\fl}^{+}=\Delta^{+} \cap \Delta_{\fl}$.
Finally, we define 
$\Delta_{\fn}:= \Delta^{+} \backslash \Delta_{\fl}\subset \Delta^{+}$.

Next, for a root $\beta \in \mathfrak{h}^*$ the co-root $\beta^{\vee} \in \mathfrak{h}$
is defined by the condition
\beq
\langle \lambda, \beta^{\vee} \rangle:
= 2 \frac{(\lambda, \beta)}{(\beta, \beta)} \ ,
\eeq
for all $\lambda \in \mathfrak{h}^*$,
where $(-, -)$ is the canonical non-degenerate bilinear form of $\mathfrak{g}$,
and $\langle -, - \rangle$ is the canonical pairing between elements of $\mathfrak{h}$ and $\mathfrak{h}^*$.
The symbol $\rho$ denotes the Weyl vector
\beq
\rho:=\frac{1}{2} \sum_{\alpha \in \Delta^+ } \alpha   \ .
\eeq

The power of \eqref{bosonic_determinant} is defined as the $\mu$-component of the character
$\textrm{ch}\, M_{\mathfrak{p}}(\lambda)$, the character of our parabolic Verma module $M_{\mathfrak{p}}(\lambda)$: 
\begin{align}
\textrm{ch}\, M_{\mathfrak{p}}(\lambda)=:\sum_{\mu} \textrm{ch}\, M_{\mathfrak{p}}(\lambda)^{\mu} e_{\mu} \;.
\label{e_mu_decomp}
\end{align}
where $e_{\mu}$ is a formal generator (with $\mu$ running over integral weights), whose linear combination gives a character.
This character is related to the character $\textrm{ch}\, M(\lambda)$ of the ordinary Verma module by \cite[Lemma 1]{Jantzen}\footnote{$M_{\mathfrak{p}}$ here is denoted by $M'$ in \cite{Jantzen}.}
\begin{align}
\textrm{ch}\, M_{\mathfrak{p}}(\lambda)= \sum_{w\in W_{\fl}} \textrm{det}(w)\, \textrm{ch}\, M(w\cdot \lambda) \;.
\label{chi_def}
\end{align}
Here $W_{\fl}$ is the Weyl group of the subalgebra $\fl$.

In general, the argument of $\textrm{ch}\, M_{\mathfrak{p}}$ in the determinant formula \eqref{bosonic_determinant},
namely $\lambda-n\beta$, is not a highest weight of the rotation group $\mathfrak{l}$. 
In those cases we can use the right hand side of \eqref{chi_def} as the 
definition of $\textrm{ch}\, M_{\mathfrak{p}}$.\footnote{For this reason it is mathematically more precise to use a different symbol for the right hand side of \eqref{chi_def}, and not the symbol $\textrm{ch}\, M_{\mathfrak{p}}$. Since this paper is primarily intended for applications
I chose to use the same symbol here for simplicity.} Consequently
$\textrm{ch}\, M_{\mathfrak{p}}(\lambda-n \beta)$  in general contains negative 
entries when expanded in terms of the basis $e_{\mu}$.

\bigskip

From the determinant formula \eqref{bosonic_determinant}, we can derive the 
simplicity criterion of Jantzen \cite[Satz3]{Jantzen}.
First, the parabolic Verma module clearly has no null states
if the set
\begin{align}
\Psi_{\lambda}^{+}:=\left\{
\beta \in \Delta_{\fn} \big|
\, n_{\beta}:=\langle \lambda+\rho, \beta^{\vee} \rangle \in \mathbb{Z}_{>0}
\right\}
\end{align}
is empty. If this set is not empty, then $\beta \in 
\Psi_{\lambda}^{+}$ potentially contributes to the 
zero of the determinant. However, there can still be cancellations for the powers in \eqref{bosonic_determinant}.
In order for this to happen, we need to have
the sum of the exponents to vanish:
\beq
\sum_{\beta \in \Psi_{\lambda}^{+}} \textrm{ch}\, M_{\mathfrak{p}}(\lambda-n_{\beta}\beta)^{\mu}=0
\eeq
for each $\mu$.
Since $\lambda-n_{\beta}\beta$ is a Weyl reflection $s_{\beta}$ of $\lambda$ by $\beta$
\beq
\lambda-n_{\beta}\beta
= \lambda -\langle \lambda+\rho, \beta\rangle  \beta
=: s_{\beta}\cdot \lambda 
\eeq
and since we have \eqref{e_mu_decomp},
we come to the following result \cite[Satz 3]{Jantzen}:
our parabolic Verma module is simple (i.e.\ contains no null states)
if and only if 
\begin{empheq}[box={\mybluebox[7pt]}]{equation}
\sum_{\beta \in \Psi_{\lambda}^{+}} \textrm{ch}\, M_{\mathfrak{p}}( s_{\beta}\cdot \lambda)=0 \;.
\end{empheq}
While this criterion (determining the positions of the zeros of the determinant) is sufficient for many purposes,
the determinant formula \eqref{bosonic_determinant} contains also the information of the multiplicities of the zeros.

\bigskip

There are several remarks on the formula \eqref{bosonic_determinant}.

First, let us consider the case of the ordinary Verma module.
In this case, we have
\begin{align}
\textrm{ch}\, M_{\mathfrak{p}}(\lambda)=\textrm{ch}\, M(\lambda) =\sum_{\mu} P(\mu)\, e_{\lambda-\mu}\;,
\end{align}
where the Kostant function $P(\mu)$
is defined to be the number of ways we can express $\mu$ 
as a sum of positive roots with positive coefficients.
In particular $P(\mu)$ is zero whenever the we have negative coefficients
in the expansion of $\mu$ in terms of positive roots.
This gives the formulas of \cite{Jantzen73,Shapovalov} (the 
Jantzen-Shapovalov determinant formula):
\beq
\textrm{det}\, M(\lambda)^{\mu}= 
\textrm{const.} \prod_{\beta\in {\Delta^{+}}} \prod_{n>0} 
\left(
\langle \lambda+\rho , \beta^{\vee} \rangle-n 
\right)^{P(\lambda-\mu-n \beta)_{\mu}} \ .
\label{Shapovalov_determinant}
\eeq
This formula is better-known in the literature.
We stress, however, that for general discussion of CFTs we need the more general formula of \eqref{bosonic_determinant}.

\bigskip
Second, the meaning of the abstract definition of $\Delta_{\fn}^{+}$ becomes clearer when we consider the example of the conformal algebra.
In this case, $\Delta^{+}$ is the set of the positive roots for the whole conformal algebra, and in particular contains
the roots corresponding to the momentum $P^{\mu}$ as well as the lowering operators of the rotation group ($J^{-}$ for $D=3$ in the standard notation;
let us here use the notation $\vec{J}^{-}$ for general $D$).
For an ordinary Verma module the $\mathfrak{l}$ is a Cartan subalgebra, and hence $\Delta_{\fl}$ is empty and 
$\Delta_{\fn, +}=\Delta_{+}$ is given by these roots, explaining the appearance of $\Delta^{+}$ in \eqref{Shapovalov_determinant}: $P^{\mu}$ as well as the rotation raising operators $\vec{J}^{-}$
generates descendants.

However, we know that this is not what we do in CFTs, and that  $P^{\mu}$ and $\vec{J}^{-}$ play rather different roles.
Since we are interested in the representations with a fixed spin,
any state should be annihilated by $\vec{J}^{-}$'s, if repeated sufficiently many times. The $P^{\mu}$'s, however, generates conformal descendants (derivatives of the operators in the state-operator correspondence), and repeated application of the $P^{\mu}$'s never annihilates the state
(unless we decouple a null state in the conformal descendant).
The descendants should be labeled by the roots for $P^{\mu}$ only, and 
we should mod out by the action of $\vec{J}^{-}$, which only changes the states inside the same representation of the rotation group (for $D=3$, $J^{-}$ changes eigenvalues of $J^3$ in the standard notation).
This points to the conclusion that for discussions of descendants we should really consider the roots in $\Delta_{\fn}=\Delta^{+}\backslash \Delta_{\fl}$,
explaining the appearance of the set $\Delta_{\fn}$ in \eqref{bosonic_determinant}.

The null states could still be at the location labeled by positive linear combinations of the elements of $\Delta_{\fn}$, say
at the location $\mathbb{Z}_{>0} \alpha_1+\mathbb{Z}_{>0} \alpha_2$
for two elements $\alpha_1, \alpha_2 \in \Delta_{\fn}^{+}$.
The surprise in the formula \eqref{bosonic_determinant} is that this does not happen, and 
the null states can and do indeed appear at $\mathbb{Z}_{>0} \alpha$
for a single element $\alpha\in  \Delta_{\fn}$, and not anywhere else.
It would be nice to provide simple intuitive explanation for this
remarkable fact.

\bigskip

Finally, let us make one small consistency check of \eqref{bosonic_determinant}.
Since $\lambda$ is the highest weight for a representation of $\mathfrak{p}$ (or rather $\mathfrak{l}$ for the case of the conformal algebra),
$w\cdot \lambda$ for $w\in W_{\fl}$ has a weight smaller than $\lambda$, unless $w=1$,
which gives the weight $\lambda$.
This in particular means that $\textrm{ch}\, M_{\mathfrak{p}}(\lambda)^{\mu}=0$ when $\mu>\lambda$.
This ensures that $\textrm{ch}\, M_{\mathfrak{p}}(\lambda-n\alpha)^{\lambda}=0$ for $n>0$, which makes sense since
$\mu=\lambda$ is a primary state in itself and hence should not admit any null states.

\subsection{Conjecture for Superalgebras}

It is straightforward to generalize the concept of a parabolic Verma module
to the case of Lie superalgebras. This is needed for the discussion of 
the superconformal algebras \cite{KacAdvM,Frappat:1996pb,Nahm:1977tg}.

The formula is very similar to the bosonic case, so we here comment briefly on some changes.
The superconformal algebra now contain fermionic generators of supersymmetry $Q^I_{\alpha}$,
fermionic superconformal generators $S^I_{\alpha}$, as well as bosonic R-symmetry generators $R^{IJ}$, with R-symmetry indices denoted by $I,J, \ldots$.
For spinors the details of the reality conditions of the spinors vary from dimension to dimension,
and we here schematically denoted them by $\alpha$. For example, in four spacetime dimensions ($D=4$)
$\alpha$ here in more standard notation represents both the dotted and un-dotted spinor indices, $\alpha$ and $\dot{\alpha}$.

Most of the stories works in parallel, replacing the representation of the bosonic Lie algebra by that of the super Lie algebra.
For example, the supersymmetric version of the decomposition \eqref{gdecomp} now reads
\beq
\mathfrak{l}=\langle D, J_{\mu \nu}, R^{IJ} \rangle  \ , \quad
\mathfrak{n}_{+} =\langle K_{\mu}, S^I_{\alpha} \rangle  \ , \quad
\mathfrak{n}_{-} =\langle P_{\mu}, Q^I_{\alpha} \rangle  \  .
\label{gdecomp2_susy}
\eeq
and a conformal primary is replaced by a superconformal primary, 
which is annihilated both by $K_{\mu}$ and $S^I_{\alpha}$.
For the definition of the inner product and the conjugation we add
\begin{align}
(Q^I_{\alpha})^{\dagger} = S^I_{\alpha} \ ,  \quad (R^{IJ})^{\dagger} = R^{IJ} \ .
\end{align}
The parabolic Verma module is spanned by superconformal descendants
\beq
M_{\mathfrak{p}}(\lambda)= \textrm{span} \left\{Q^{I_1}_{\alpha_1} \ldots Q^{I_l}_{\alpha_l}  P^{\mu_1} P^{\mu_2} \ldots P^{\mu_k} |\Omega_{\lambda}^a\rangle
\right\}  \ .
\eeq
From mathematical standpoint this is not the most general parabolic Verma module:
$\mathfrak{l}$ does not contain fermionic/odd elements.

\bigskip

A determinant formula for a generalized Verma module 
for a superalgebra does not seem to be 
known in the literature, at least in the explicit form as stated above for the bosonic case\footnote{See however \cite{osp_1,osp_2}
for recent discussions on orthosymplectic Lie superalgebras}.
The formula question should be a natural generalization of \eqref{bosonic_determinant}.
Our conjecture for the determinant formula, for the case where $\mathfrak{l}$ does not contain fermionic generators, 
and for $\mathfrak{g}$ being one of the finite-dimensional superconformal algebras classified in \cite{Nahm:1977tg},
is 
\begin{align}
\begin{split}
\textrm{det}\, M(\lambda)^{\mu}& = 
\textrm{const.} 
 \prod_{\beta\in {\overline{\Delta}_{\fn, 0}}} \prod_{n>0} 
\left(
\langle \lambda+\rho , \beta^\vee \rangle- \frac{n}{2} 
\right)^{\textrm{ch}\, M_{\mathfrak{p}}(\lambda-n \beta)_{\mu}}
 \\
&\times  \prod_{\beta \in \Delta_{1}^{+},\, (\beta, \beta) \ne 0} 
\prod_{n>0}
\left(
\langle \lambda+\rho , \beta^\vee \rangle
-\frac{2n-1}{2}\right)^{\textrm{ch}\, M_{\mathfrak{p}, \beta}(\lambda-(2n-1)\beta)_{\mu}}\\
&\times \prod_{\alpha \in \Delta_{1}^{+},\,  ( \beta, \beta )=0} 
\left(
\lambda+\rho , \beta
\right)^{\textrm{ch}\, M_{\mathfrak{p}}(\lambda-\beta)_{\mu}} 
 \ .
\label{fermionic_determinant}
\end{split}
\end{align}
Here we denoted by $\Delta^+$ the set of positive roots, 
which are decomposed into the set of positive bosonic roots 
$\Delta_{0}^+$ and that of positive fermionic roots 
$\Delta_{1}^+$: $\Delta^+=\Delta_{0}^+ \cup \Delta_{1}^+$.
Otherwise $\Delta_{0, \fn}$ is defined in a similar manner to 
$\Delta_{\fn}$ for the bosonic case.
We then define its subspace
$\overline{\Delta}_{0, \fn}$ by
\begin{align}
\overline{\Delta}_{0, \fn}: =\left\{
\beta\in \Delta_{0, \fn}\ , \quad\beta/2\notin \Delta_{1}
\right\}
\end{align}
In the determinant formula the product is over the set $\overline{\Delta}_{0, \fn}$,
and not over the whole of $\Delta_{0, \fn}$.
This is because for a fermionic root $\beta \in \Delta_{1}$ with  $(\beta, \beta) \ne 0$,
$2\beta$ is a bosonic root, and we need to avoid over-counting.

We modified the definition of the Weyl vector to be
\beq
\rho=\frac{1}{2} \left(
\sum_{\alpha \in \Delta^+_{0}} \alpha  
-
\sum_{\alpha \in \Delta^-_{1} }\alpha  
\right) \ ,
\eeq
The character $\textrm{ch}\, M_{\mathfrak{p}, \beta}$ is 
$\textrm{ch}\, M_{\mathfrak{p}}$ with the contribution from the fermionic root $\beta$ removed.
More precisely, the general definition, generalizing 
\eqref{chi_def}, is given by
\begin{align}
\textrm{ch}\, M_{\mathfrak{p}, \beta}(\lambda):= \sum_{w\in W_{\fl}} \textrm{det}(w)\, \textrm{ch}\, M_{w\cdot \beta} (w\cdot \lambda) \;.
\end{align}
with $\textrm{ch}\, M_{\beta}(\lambda)$ being the (ordinary) Verma module character with 
contributions from the fermionic root $\beta$ removed:
\begin{align}
\textrm{ch}\, M_{\beta}(\lambda):= \textrm{ch}\, M (\lambda) \, (1+e_{-\beta})\;.
\end{align}

One justification for our conjecture is that
for a Verma module this formula reduces to the formula by Kac \cite{Kac_1986}\footnote{The 
determinant formula for a (non-parabolic, i.e.\ ordinary) Verma module, which was first
written down by Kac in \cite{Kac_LNM,Kac_LNP}, turned out to be incorrect. Kac himself later
corrected his formula in \cite{Kac_1986}.
}.
We can also check that this formula reduces to the 
previous formula \eqref{bosonic_determinant} for a bosonic Lie algebra.\footnote{Note added on revision: this conjecture was subsequently proven by Y.~Oshima and the author \cite{Oshima:2016gqy}.}

In \eqref{fermionic_determinant}, the fermionic roots play rather different roles depending on whether or not the root has 
zero norm or not. 

\section{Remarks on Applications}\label{sec.applications}

One application of the determinant formula is that it places severe constraints on the possible singularities of the 
conformal blocks, as functions of intermediate operator dimensions.

For concreteness let us consider a 4-point functions of 
four scalar operators $\mathcal{O}_1(x_1), \ldots, \mathcal{O}_4(x_4)$
with operator dimensions $\Delta_1, \ldots, \Delta_4$.
Conformal symmetry constraints the 4-point function to be of the form
\beq
&\langle \mathcal{O}_1(x_1) \mathcal{O}_2(x_2) \mathcal{O}_3(x_3) \mathcal{O}_4(x_4) \rangle \\
&\qquad= \sum_{\mathcal{O}\in \mathcal{O}_1\times \mathcal{O}_2}
\lambda_{\mathcal{O}\,\mathcal{O}_1\mathcal{O}_2}^2
\frac{1}{x_{12}^{2(\Delta_1+\Delta_2-\Delta)}x_{34}^{2(\Delta_3+\Delta_4-\Delta)}}\, \mathcal{G}_{\Delta,\vec{l}}(u,v) \;,
\label{blockdef}
\eeq
where $x_{ij}:= x_i-x_j$, $\lambda_{\mathcal{O}\,\mathcal{O}_1\mathcal{O}_2}$ are OPE structure coefficients,
and $u, v$ are the conformal cross-ratios defined by
\begin{align}
u:=\frac{x_{12}^2 x_{34}^2}{x_{13}^2 x_{24}^2}\;, \quad
v:=\frac{x_{14}^2 x_{23}^2}{x_{13}^2 x_{24}^2} \;.
\end{align}
The sum in \eqref{blockdef} is over all the conformal primaries $\mathcal{O}$
(with dimension $\Delta$ and spin $\vec{l}$) in the OPE of the operators 
$\mathcal{O}_1$ and $\mathcal{O}_2$.
The function $\mathcal{G}_{\Delta,\vec{l}}(u,v)$ is the conformal block for the scalar four-point function,
and once we know this function we could start exploring the constraints from
crossing symmetry and unitarity, as has been done recently in the literature.

We are interested in the poles of the conformal blocks
as a function of the operator dimension $\Delta$.
The fact that the conformal block, and hence the 4-point function,
has poles can be seen from the following representation of the 4-point function
\beq
&\langle \mathcal{O}_1(x_1) \mathcal{O}_2(x_2) \mathcal{O}_3(x_3) \mathcal{O}_4(x_4) \rangle \\
&=\sum_{\alpha=\mathcal{O}, P^{\mu}\mathcal{O}, \ldots ,  \, ; \, \mathcal{O}\in \mathcal{O}_1 \times \mathcal{O}_2}
\frac{\langle \mathcal{O}_1(x_1) \mathcal{O}_2(x_2) | \alpha \rangle
 \langle \alpha | \mathcal{O}_3(x_3) \mathcal{O}_4(x_4) \rangle
 }{\langle \alpha | \alpha \rangle}   \ , 
 \label{alpha_decomp}
\eeq
where the sum is now over all the descendants $\alpha=P^{\mu_1} \cdots P^{\mu_n} \mathcal{O}$
of the conformal primary  $\mathcal{O}$. This equation represents the fact that the set of conformal primaries ($\mathcal{O}$'s), as well as their descendants, span a complete basis of states. This is the case for a generic value of $\Delta$.

We find from \eqref{alpha_decomp} that the divergence can arise when the norm of $\alpha$ is zero, namely when $\alpha$ is a null state (singular vector).\footnote{As in \eqref{blockdef}, the four-point function diverges in the OPE limit where a pair of $x_i$'s coincide. Here we are discussing divergence which arise with fixed values of the coordinates $x_i$'s.} This happens precisely when the value of the operator dimension $\Delta$ is fine-tuned (to a value, say, $\Delta=\Delta_{\star}$) such that one of the determinants (for a generalized Verma module determined for given $\Delta$ and $\vec{l}$) vanishes (for some value of $\mu$).
Note that in general such singularities are generically order one poles, but in general can be higher orders poles.

For such fine-tuned values of $\Delta$, the presence of the null states means that the parabolic Verma module is 
reducible as a representation of the (super)conformal algebra, and that it decomposes into the null descendants, and the rest with the null descendants decoupled (namely, the quotient of the parabolic Verma module modulo null descendants). 
This fact has a counterpart in the conformal blocks. The conformal block at $\Delta$ generic is singular at $\Delta=\Delta_{\star}$ and is not smoothly connected with that for $\Delta=\Delta_{\star}$, which is defined from \eqref{alpha_decomp} with null descendants removed from $\alpha$. The singular behavior as $\Delta\to \Delta_{\star}$
is governed by null descendants, and this fact is the basis for the recursion relation of \cite{Penedones:2015aga}.

As the discussions above makes clear, we reach the same conclusion if we 
have 4-point functions of more general operators with arbitrary spins.
This places a rather stringent constraint on the analytic behavior of the conformal blocks,
we can derive this constraint purely from the representation-theoretic analysis.

One cautionary remark is that 
the inverse of the preceding statement does not hold: namely, even when $\Delta$ is chosen such that the 
determinant formula vanishes, the conformal block might not have a singularity in general.
For example, the coefficient coming from the three-point function (i.e.\ residue)
could still be zero, for example by representation-theoretic reasons.
A good example for this is the discussion of the scalar conformal blocks in \cite{Penedones:2015aga},
where the poles for the type IV states are absent from the conformal block for the four scalar operators.
Even when it has a singularity, we could have double and higher-order poles. 
Indeed, the four-dimensional scalar conformal block
discussed in \cite{Dolan:2000ut} does contain such double poles.

Once we know the positions of poles of conformal blocks, we can also 
try to study the residues of the conformal blocks.
As illustrated in \cite{Penedones:2015aga},
this requires an explicit form of the null states, about which little seems to be known at present (except for the cases studied in \cite{Penedones:2015aga}).
Once we know the poles and residues we can in addition analyze the behavior $\Delta\to \infty$, to 
derive recursion relations for conformal blocks \cite{Penedones:2015aga}\footnote{The recursion relation for the scalar block
was first obtained in \cite{Kos:2013tga}. See also \cite{Kos:2014bka,Iliesiu:2015akf} for recent related work}.
This procedure, however, could in general be complicated by the presence of double poles and higher order poles.

We also note that the determinant formula is useful for the systematic derivations of the unitarity bounds (see \cite{Penedones:2015aga}). The constraint of the unitarity is that the norm of the all states are positive, and when this fails as we change the parameters
some of the states will necessarily have zero norm. In other words, we can derive unitarity bound from the absence of the null states.

In the literature we often derive the unitarity bound by working out the absence of the null states 
in the first level descendant of the primary. However, for a complete derivation of the unitarity bound we need to 
make sure that no stronger constraints arise from the further descendants,
and it is rather non-trivial to show that this is indeed the case\footnote{See \cite{Evans} for $D=3$ and \cite{Mack:1975je} for $D=4$, see also \cite{Grinstein:2008qk} for more recent related work. However, the proofs for $D=3, 4$ are rather involved, and do not generalize easily for general $D$.
Our determinant formula, by contrast, is applicable to general $D$.}.
Note that the classic paper of \cite{Minwalla:1997ka} discuss the unitarity bound using the Kac's criterion \cite{Kac_LNM}.
However, the representations discussed in \cite{Kac_LNM} are the (non-parabolic, i.e.\ ordinary) Verma modules for the Cartan subalgebra,
and as we have seen what is relevant for our discussion of conformal blocks is a more general 
parabolic Verma module.

Finally, instead of avoiding the null states one can try to taking advantage of them. Just as in the case of two dimensions, we might 
eliminate the null states and define a higher-dimensional counterparts of the minimal models.


\section*{Acknowledgments}
The author would like to thank in particular 
J.\ Penedones and E.\ Trevisani for related collaboration \cite{Penedones:2015aga},
without which this paper would never have been materialized.
He would also like to thank Y.\ Oshima for discussion on mathematical aspects
and also L.\ Dolan, 
S.\ Giombi, J.\ Humphreys, J.\ Maldacena, D.\ Simmons-Duffin, E.\ Witten
for stimulating discussion.

This work was close to completion in late 2013, however
was kept in the author's desk for more than two years.
I would like to thank V.\ S.\ Rychkov for encouraging me to complete this work.

This research is supported by WPI program (MEXT, Japan), by JSPS Program for Advancing Strategic International Networks to Accelerate the Circulation of Talented Researchers, by JSPS KAKENHI Grant No.\ 15K17634, and by Adler Family Fund. 
He would also like to thank 
KITP UCSB (``New Methods in Nonperturbative Quantum Field Theory'', Grant No.\ NSF PHY11-25915) for hospitality where part of this work has been performed. 
The contents of this paper was presented at the conference 
``Back to the Bootstrap IV'', University of Porto, July 2014.


\bibliographystyle{nb}
\bibliography{bootstrap}

\end{document}